\documentclass[aps,pre,twocolumn,showkeys,superscriptaddress,groupedaddress]{revtex4}
\usepackage{graphicx}
\usepackage{epsfig}
\usepackage{amssymb}
\usepackage{amsmath}
\usepackage{color}
\begin{document}

\title{A logistic model for flowing particles}

\author{Byung Mook \surname{Weon}}
\email{bmweon@skku.edu}
\affiliation{Soft Matter Physics Laboratory, School of Advanced Materials Science and Engineering, SKKU Advanced Institute of Nanotechnology (SAINT), Sungkyunkwan University, Suwon 16419, South Korea}
\affiliation{Research Center for Advanced Materials Technology, Sungkyunkwan University, Suwon 16419, South Korea}
\affiliation{Department of Biomedical Engineering, Johns Hopkins University, Baltimore, Maryland 21218, USA}

\date{\today}

\begin{abstract}
Counting how many particles pass through a specific space within a specific time is an interesting question in applied physics and social science. Here a logistic model is developed to estimate the total number of flowing particles. This model sheds light on a collective contribution of particle growth rate and transient probability within a specific space in particle counting. This model may offer a basic concept to understand transport dynamics of flowing particles.
\end{abstract}


\maketitle

How many particles have passed there? This question is simple but significant in many physical, biological, and social situations \cite{Watson,Jin,Marchetti}. Counting the total number of flowing particles is often a difficult task because of complexity in particle mobility and transport dynamics. Conceptually, this question is similar to a population dynamics that is controlled by birth and death rates or immigration and emigration \cite{Marchetti}. In mathematical biology, the simplest population growth model is the Malthusian exponential model where the total population increases exponentially with time \cite{Stokes}. The logistic model is widely established in many fields for modeling and forecasting population \cite{Verhulst}. The logistic growth dynamics assumes that the total population grows exponentially and saturates to an upper limit, producing a typical $S$-shaped curve. The upper limit represents a capacity limit in the system. In a confined space, there may be a capacity limit and thus the logistic model would be appropriate in particle counting.

In this article, the logistic model is developed to understand flowing particles and particle counting. This model sheds light on a collective contribution of particle mobility and growth rate to the total number of particles. This model is applicable for both of static and mobile particles, probably offering a new framework for understanding transport dynamics of static or mobile particles.

First, consider a physical situation for flowing particles, where a fixed number of flowing particles occupy a limited number of positions in a space, as illustrated in {\bf Fig. 1}. As flowing particles move through a space together like flowing crowds \cite{Low,Ouellette,Helbing,Karamouzas}, the total number of particles initially increases with time, reach a peak for a while, and eventually diminishes with time. In this situation, the number of particles can be modeled by a combination of particle growth and decay dynamics. This physical situation can be modeled with the factors of the first (or final) particle contribution $a$, the rate of growth (or decay) $b$, and the maximum capacity in the place $c$ (physically, $c$ is set by a multiple of the occupation space $\alpha$ and the population density $\beta$ as $c = \alpha\beta$).

\begin{figure}[t]
\includegraphics[width=9cm]{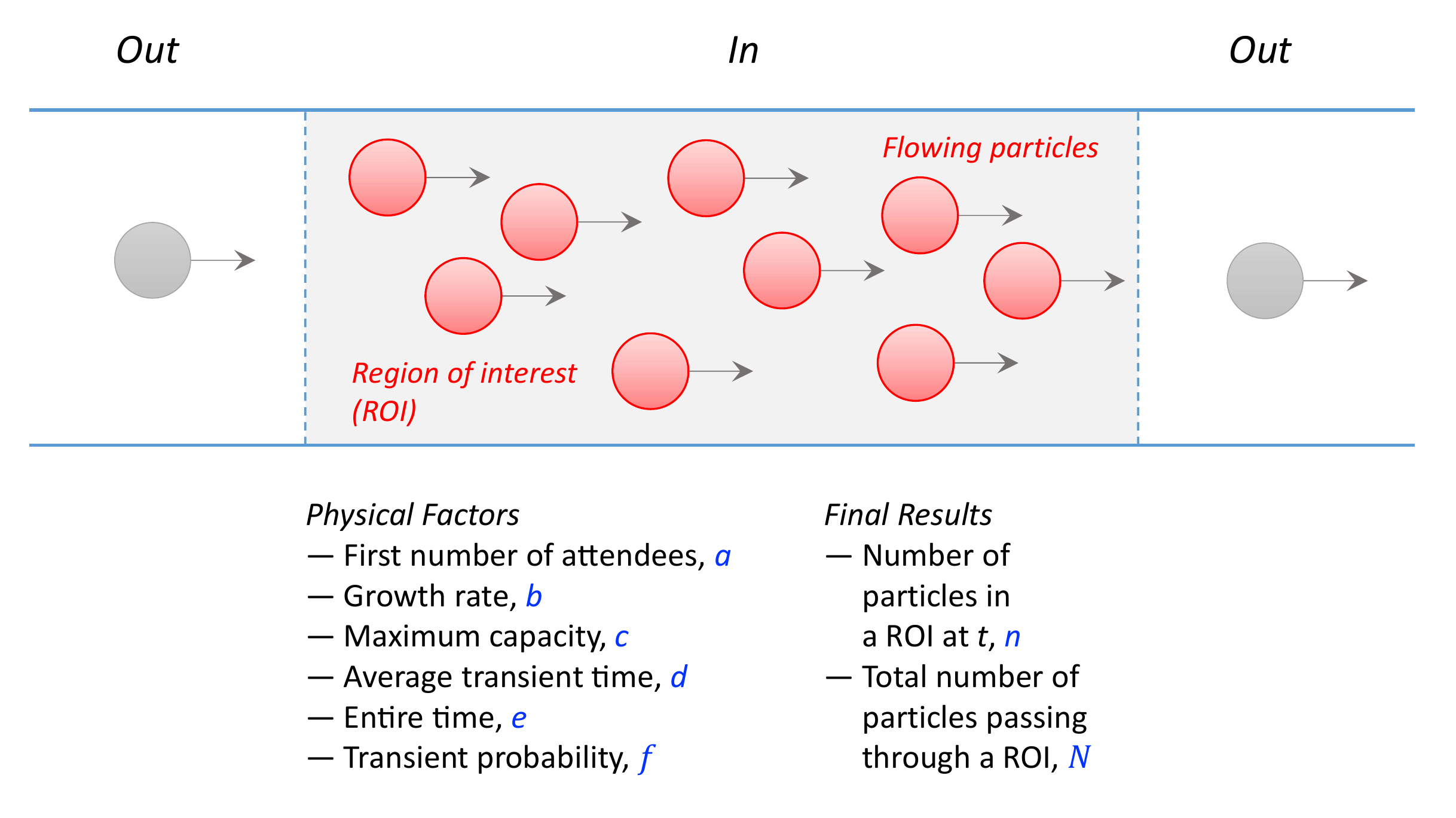}
\caption{Illustration of a situation: when particles are flowing in a region of interest (ROI, grey) and their physical factors are given ($a-f$), an important question is how many particles ($n$ or $N$) have passed through the ROI during a period.}
\label{fig:1}
\end{figure}

Next, to quantify the hydrodynamic aspects of flowing particles \cite{Bain,Hughes}, the average transient time $d$ is considered as follows. The transient time is the spent duration for particles to stay by occupying the limited positions and is responsible for the particle mobility. Assuming the entire time $e$ for growth and decay, the transient probability $f$ is calculated as $f = d/e$. By taking the transient probability, the particle mobility can be quantified.

The transient probability is useful to characterize the nature of static or mobile particles. For instance, let's think about the following two situations. In the first case, most particles may stay to pass through for a while (e.g., for 30 minutes) during the entire time (e.g., for 2 hours), suggesting the transient probability to be $f = \frac{30}{120} = 0.25$ on average. In the second case, most particles may stay for a while (e.g., for 110 minutes) during the entire time (e.g., for 2 hours), indicating $f = \frac{110}{120} = 0.92$. The first case corresponds to {\it mobile} particles ($f \ll 1.0$), while the second case to {\it static} particles ($f \approx 1.0$).

\begin{figure}[t]
\includegraphics[width=9cm]{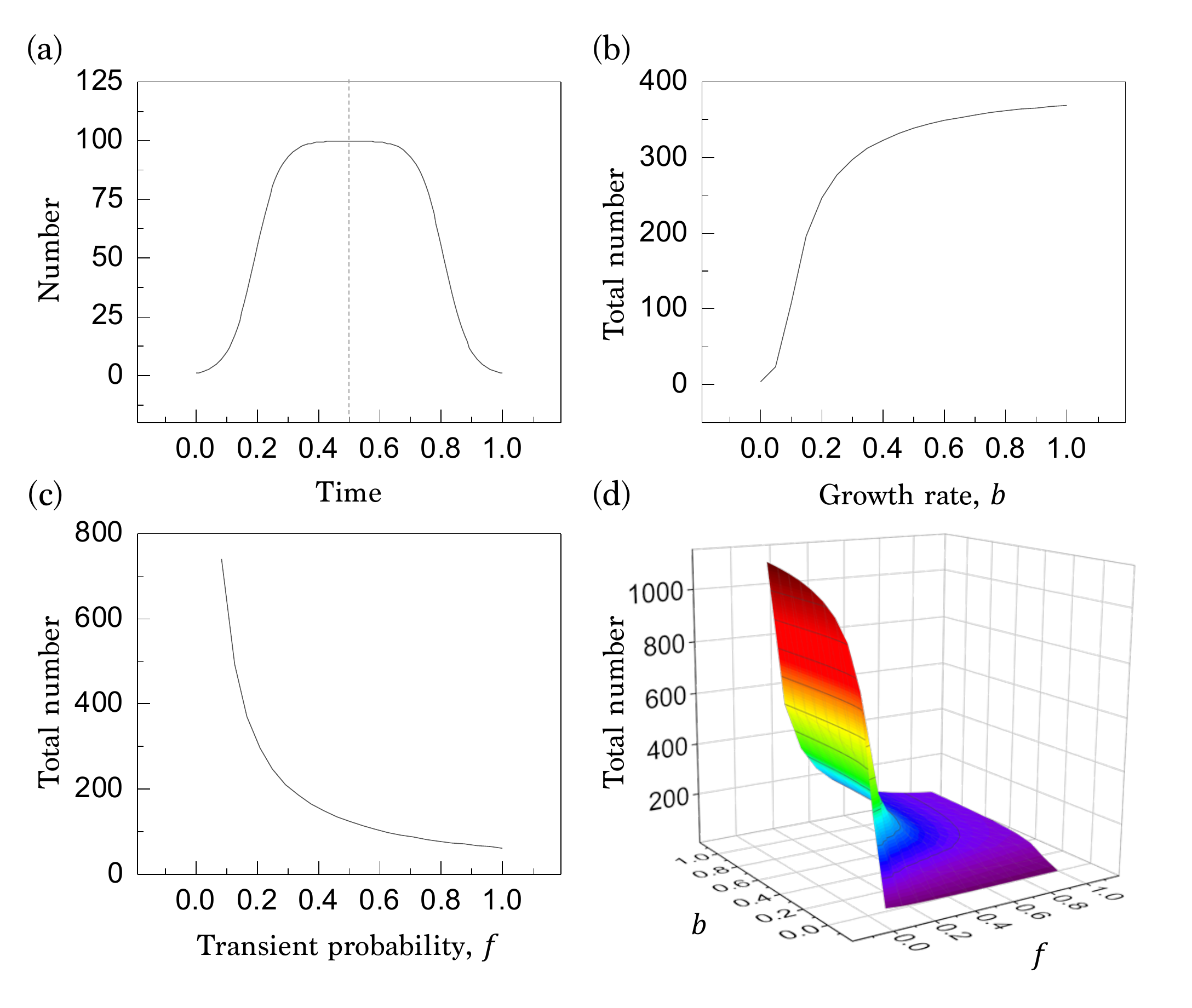}
\caption{The logistic model: (a) the particle number $n(t)$ changes with time [$a = 1.0$, $b = 0.2$, $c = 100$, $d = 30$, and $e = 120$], (b) by the contribution of the growth rate $b$ [by fixing $a = 1.0$, $c = 100$, $d = 30$, and $e = 120$], (c) by the contribution of the transient probability $f$ [by fixing $a = 1.0$, $b = 0.2$, $c = 100$, and $e = 120$], and (d) by the collective contribution of the growth rate $b$ and the transient probability $f$ [by fixing $a = 1.0$, $c = 100$, and $e = 120$].}
\label{fig:2}
\end{figure}

To describe static or mobile particles with the logistic model, the logistic growth dynamics is applied prior to a peak as \cite{Verhulst,Stokes,Jin}:
\begin{equation}
  n(t) = \frac{ac}{a+(c-a)e^{-bt}},
\end{equation}
and after passing a peak, the logistic decay dynamics is applied as:
\begin{equation}
  n(t) = \frac{ac}{a+(c-a)e^{-b(e-t)}}.
\end{equation}
Here $n(t)$ is the number of particles at a moment and is determined by the first (or final) number of particles $a$, the growth (or decay) rate $b$, the maximum capacity $c$, the average transient time $d$, the entire time $e$, the transient probability $f = d/e$, and the peak time $g = \frac{1}{2}e$. By integrating $n(t)$ with respect to $t$ and dividing it by the average transient time, the total number $N$ can be estimated as:
\begin{equation}
  N = \frac{\int n(t)\;\mathrm{d}t}{d}.
\end{equation}

\begin{figure}[t]
\includegraphics[width=9cm]{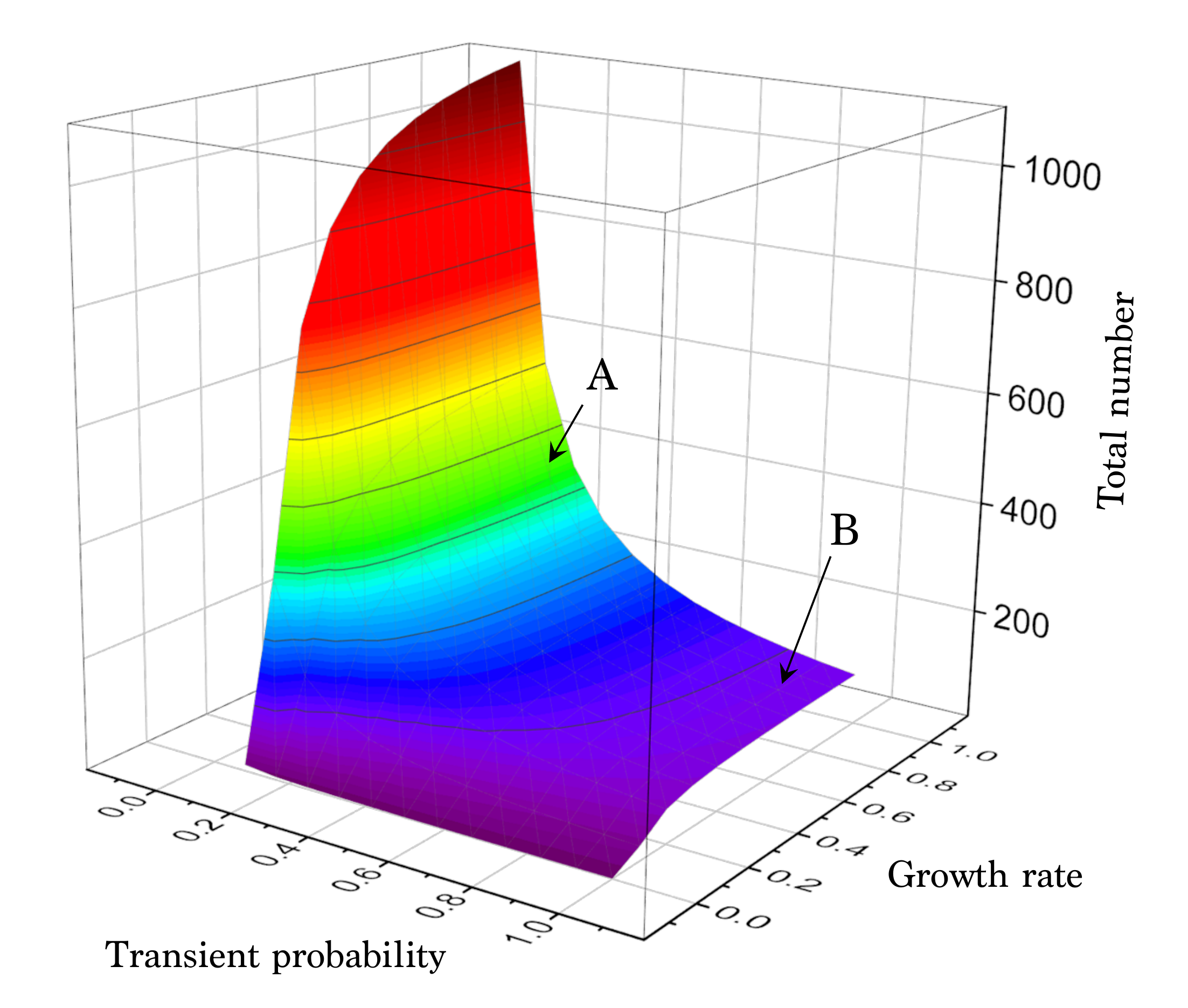}
\caption{The characterization of static and mobile particles: the total number of particles is determined by the collective contribution of the transient probability and the growth rate. For instance, the total number increases by 3.7 times when the transient probability decreases to $f = 0.25$ (mobile particles, marked A) from $f = 0.95$ (static particles, marked B) for the same growth rate $b = 1.0$ [by fixing $a = 1.0$, $c = 100$, and $e = 120$].}
\label{fig:3}
\end{figure}

As demonstrated in {\bf Fig. 2}, the logistic model is appropriate to evaluate how the particle number changes with time by the physical factors in the logistic model. In {\bf Fig. 2(a)}, for the physically feasible conditions, $a = 1.0$, $b = 0.2$, $c = 100$, $d = 30$, and $e = 120$ are assumed (here, time is normalized). Controlling the factors, the particle number for static or mobile particles is counted during particle growth [{\bf Eq. (1)}] and decay dynamics [{\bf Eq. (2)}]. For simplicity, the growth dynamics is assumed to be symmetric with the decay dynamics. In {\bf Fig. 2(b)}, the contribution of the growth rate $b$ is tested by fixing the other conditions in {\bf Fig. 2(a)} except for the variable $b$ [$a = 1.0$, $c = 100$, $d = 30$, and $e = 120$]. Interestingly, the total number significantly increases with the growth rate $b$. In {\bf Fig. 2(c)}, the contribution of the transient probability $f$ is tested by fixing the other conditions in {\bf Fig. 2(a)} except for the variable $f$ [$a = 1.0$, $b = 0.2$, $c = 100$, and $e = 120$]. Interestingly, the total number is inversely proportional to the transient probability $f$. The collective contribution of the growth rate and the transient probability is illustrated in {\bf Fig. 2(d)} [by fixing $a = 1.0$, $c = 100$, and $e = 120$], showing that the total number is significantly affected by the transient probability for most $b$ values ($b \gtrsim 0.2$); that is, the particle mobility is crucial to determine the total number of flowing particles.

The logistic model is appropriate to characterize the nature of static or mobile particles. The total number of particles is illustrated in {\bf Fig. 3} as a function of the transient probability and the growth rate [by fixing $a = 1.0$, $c = 100$, and $e = 120$]. Most interestingly, the total number is significantly affected by the transient probability, rather than the growth rate. In particular, the total number significantly increases by 3.7 times when the transient probability decreases to $f = 0.25$ ($N = 3.7c$ as marked A) from $f = 0.95$ ($N = 0.97c$ as marked B) for the same growth rate $b = 1.0$. This result clearly shows why mobile particles are more than static particles. It is noteworthy that the logistic model is applicable for both static and mobile particles by simply adjusting the physical factors. To generalize the result, the total number of particles becomes more than the maximum capacity for mobile particles ($N \gg c$) and becomes less than or equal to the maximum capacity for static particles ($N \leqslant c$).

\begin{figure}[t]
\includegraphics[width=9cm]{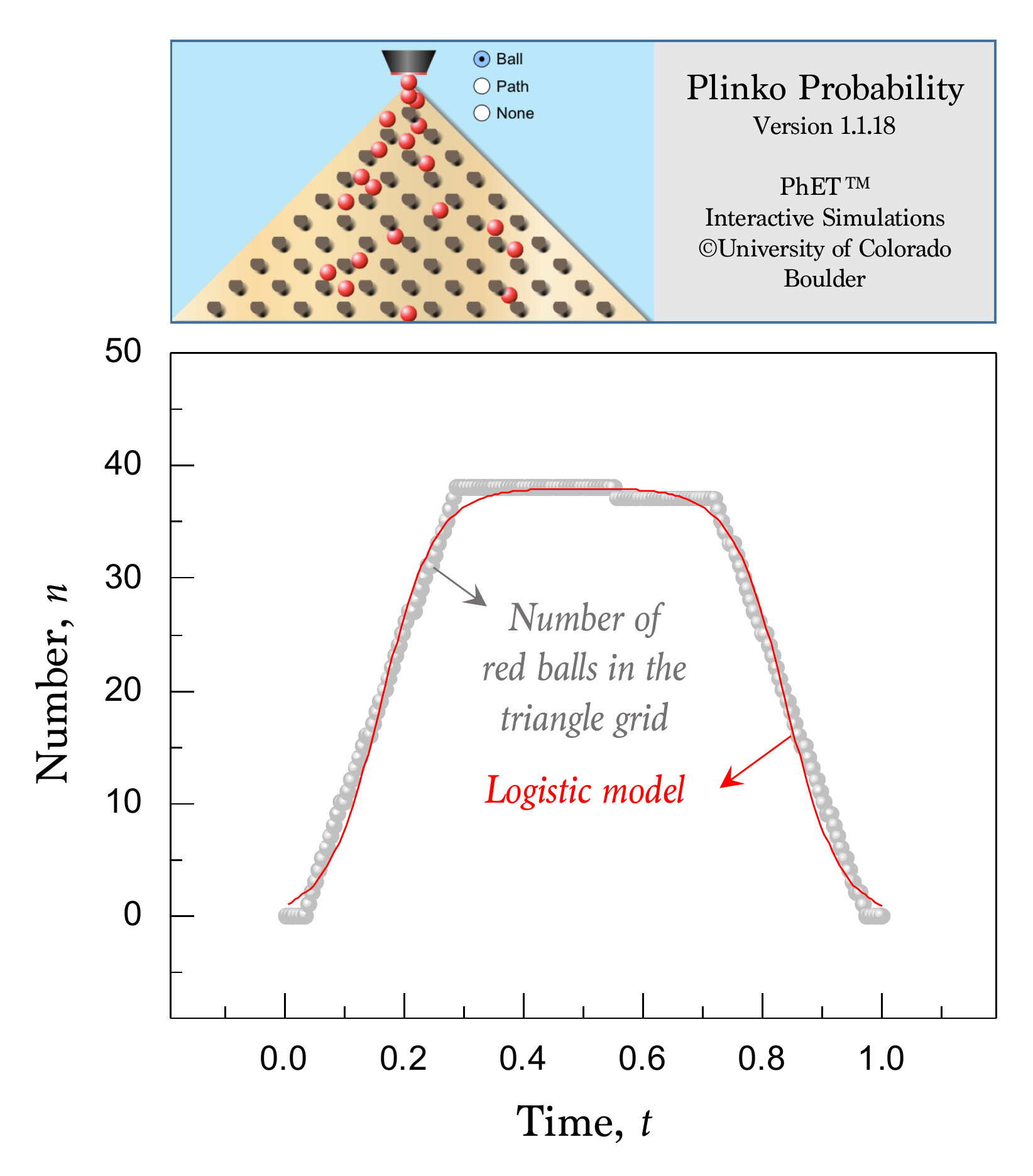}
\caption{Simulation of falling balls through a triangle grid of peds for the Plinko Probability by the PhET interactive simulations (Movie 1). The number of red balls in the triangle grid increase with time at $t < g$ and decrease with time at $t > g$, showing a good agreement with the logistic model with $a = 1.0$, $b = 0.135$, $c = 38$, $d = 42.2$, and $e = 165$ ($N/c = 2.6$).}
\label{fig:4}
\end{figure}

To demonstrate the validity of the logistic model, a simulation of falling balls through a triangle grid of peds was tested with help of the Physics Education Technology (PhET) interactive simulations (https://phet.colorado.edu) \cite{Perkins,Wieman}. In {\bf Fig. 4} (see {\bf Movie 1}), the number of red balls in the triangle grid increase with time at $t < g$ and decrease with at $t > g$. The measured ball number is compared with the logistic model with $a = 1.0$, $b = 0.135$, $c = 38$, $d = 42.2$, and $e = 165$ ($N/c = 2.6$), providing a good agreement between simulation and model.

Counting the total number of particles, both {\it a priori} and in real-time, can be applied for human crowds and planning crowd safety in places of public assembly \cite{Botta,Henke,Still}. In principle, human crowds are likely to stay or move in a place like flowing particles \cite{Hughes}. Direct counting methods are not available despite many modern technologies with artificial intelligence, drone, or visual analysis \cite{Botta,Henke}. Probably, the logistic model for flowing particles would be applied to estimate the total number of crowds. The logistic model for particle counting would be broadly applicable to estimate the particle transport through porous media in applied physics, the total number of clients visiting a store in economics \cite{Henke}, the crowd size of a protest in sociology \cite{Still}, and the growth dynamics of bacteria in a specific colony in biology \cite{Sibilo}. Further studies are required to verify the applicability of the logistic model in a variety of systems.

In conclusion, this study shows that the logistic model is appropriate to estimate the total number of flowing particles and is available for both static and mobile particles. The numerical demonstration of the logistic model clearly shows how the particle number changes with time according to the particle mobility and the growth dynamics. Practically, in physical, social, or ecological situations, the logistic model is applicable by identifying the transient probability and the growth rate.

{\bf Acknowledgments.}
This research was supported by Basic Science Research Program through the National Research Foundation of Korea (NRF) funded by the Ministry of Education (NRF-2016R1D1A1B01007133, 2019R1A6A1A03033215) and also supported by the Korea Evaluation Institute of Industrial Technology funded by the Ministry of Trade, Industry and Energy (20000423, Developing core technology of materials and processes for control of rheological properties of nanoink for printed electronics).

\end{document}